# Evaluating Malware Forensics Tools

Ian M. Kennedy[*,1], Arosha K. Bandara[1], Blaine Price[1]

[*] All correspondence should be sent to: ian.m.kennedy@open.ac.uk
[1] School of Computing & Communications, The Open University, Milton Keynes, United Kingdom

## Abstract

We present an example implementation of the previously published Malware Analysis Tool Evaluation Framework (MATEF) to explore if a systematic basis for trusted practice can be established for evaluating malware artefact detection tools used within a forensic investigation. The application of the framework is demonstrated through a case study which presents the design of two example experiments that consider the hypotheses: (1) Is there an optimal length of time in which to execution malware for analysis and (2) Is there any observable difference between tools when observing malware behaviour? The experiments used a sample of 4,800 files known to produce network artefacts. These were selected at random from a library of over 350,000 malware binaries. The tools *Process Monitor* and *TCPVCon*, popular in the digital forensic community, are chosen as the subjects for investigating these two questions. The results indicate that it is possible to use the MATEF to identify an optimal execution time for a software tool used to monitor activity generated by malware.

## Keywords
Malware forensics; Tool evaluation; Trusted practice; Malware; Expert evidence

## 1. Introduction and background

The role of expert digital forensic evidence in a courtroom has and continues to go through a transformative change in recent years. In England and Wales, digital evidence was originally inadmissible without supporting evidence to attest to its reliability and was later bestowed with a presumption of being valid, 'until evidence to the contrary is led' (Lloyd, 2020). In other words, digital evidence tendered to a court in England and Wales was trusted to be reliable. The notion of trust is succinctly encapsulated by Duranti and Rogers (2012) as "willingly acting without the full knowledge needed to act." Due to its lay background, the courts have been obliged to place some degree of trust in expert evidence in terms of both the evidence itself and the expert producing it for court.

However, the landscape today for expert evidence in general is quite different. Scientific evidence tendered to a court in the USA has for some time now been required (depending on the state) to meet the Daubert Standard, set by the judge in *Daubert v Merrell Dow Pharmaceuticals Inc.* (1993). The standard tightens the admissibility criteria to assist judges in assessing the credibility of scientific testimony. In England and Wales the Codes of Practice and Conduct (the "Codes") of the Forensic Science Regulator (FSR) obligate (but currently do not mandate) that practitioners demonstrate greater levels of scientific practice in their approach to the production of evidence. The remaining jurisdictions of the UK (Scotland and Northern Ireland) both operate to the same international standard (ISO, 2005) as the Codes, which apply to England and Wales (Forensic Science Regulator, 2021a).

Malware forensics a field that is surfacing within universities as a whole course/module (University of Portsmouth, 2019) or as part of related modules, such as Digital Forensics (University of London, 2020). Similar to the early days of computer forensics, there are (at the time of writing) no tools that are dedicated to malware forensics, in the same way there are now dedicated mobile forensics or digital forensics tools. Hence, the discipline is typically practiced using tools capable of only rudimentary malware analysis. Used by digital forensic investigators, rather than specialist malware analysts, these tools are typically used to answer investigative questions that are much more limited in scope (e.g.: to investigate the veracity of a Trojan defence) than the more comprehensive malware analysis piece a malware analyst may undertake.

Furthermore, there is little published material establishing a formal or otherwise scientific basis for procedures applied to conducting a malware forensic investigation; and more specifically for evaluating the tools to do so. Hughes and Karabiyik (2020) report that "entire domains of forensic tool testing", such as "malware analysis, remain unspecified." It is of little surprise then, that the validity of using malware scanners as part of an investigation is also open to challenge (Hughes and Varol, 2020).

In response to this gap, we pose the question, "Can a systematic basis for trusted practice be established for evaluating malware artefact detection tools used within a forensic investigation?" In our previous work (Kennedy et al., 2020) we addressed this question and identified a set of requirements that could be addressed by the Malware Analysis Tool Evaluation Framework (MATEF). Building on this, the contributions of this paper are

(a) an illustration of how the components of the MATEF can be deployed to address different types of research questions related to malware analysis tool evaluation;
(b) the demonstration of the MATEF can be instantiated in an automated virtualised environment to consider two example questions, posed to evaluate a malware analysis tool.

Thus, the MATEF is not a malware analysis tool, but rather an environment in which to evaluate the tools used to perform malware forensic analysis.

The structure of this paper is as follows: §2 explores related work; §3 presents the research design; §4 provides an example implementation of the MATEF; §5 offers a case study exploring the impact of execution time when analysing malware; finally, §6 draws conclusions and suggests further work.

## 2. Related work

The evaluation of software has long been established in software engineering through the process of Validation and Verification (Boehm, 1989). These qualities are defined in terms of being 'fit for purpose' by the FSR in their Codes (Forensic Science Regulator, 2021b). The Codes are based upon the ISO/IEC 17025 standard (ISO, 2005) and the definitions contained therein. However, there is room for interpretation in these definitions, as the expression 'fit for purpose' is not defined, nor is the threshold at which a 'method, process or device' becomes 'fit'. Also not included is any indication of which measures (metrics) are to be used or their weighting on the outcome that determines their 'fitness for purpose'.

Ayers (2009) provides seven metrics to measure the "efficacy and performance" of digital forensic tools. However, there are ambiguities in the definitions provided, such as how a tool is considered to "fail during an investigation".

In addition, clarity on expectations of tools and stakeholders is not fully established or understood by the community.  For example, Bhat et al. (2020) conclude from a study of four leading computer forensic tools that each tool "failed to identify and combat" anti-forensic attacks, such as data wiping. However, this is not a failing of a tool if the data itself is absent. Similarly, a difference of opinion exists regarding what requirements should be specified for accreditation purposes. Marshall and Paige (2018) argue that the FSR's Codes (linked to ISO 17025) lack technical requirements and instead focus on customer requirements. This is in contrast to the Tully et al. (Tully et al., 2020) who state the onus is on practitioners to have "clearly specified" these tool and method requirements.

Elsewhere, a variety of evaluation criteria is offered in a handful of guidelines and best practice projects, such as The Computer Forensic Tool Testing (CFTT) project at the National Institute for Science and Technology (NIST) (NIST, 2019). Considered by some to be rigorous (Liang, 2010), criticisms of the project include that published report relate to older versions of current software tools and that it is largely focused on acquisition (Guo and Slay, 2010), (Newsham et al., 2007) and Sommer (2010) who points out the tests completed by CFTT are but a "tiny subset" of the functionality that needs to be tested.

Unlike NIST who developed specifications, plans and assertions, the Scientific Working Group on Digital Evidence (SWGDE) (SWGDE, 2021a) have developed a more relaxed approach to forensic tool testing by producing test guidelines and templates (SWGDE, 2021b) since their inception in 1998.  A similar approach has been adopted by the Department of Defense Cyber Crime Center (DC3) (DC3, 2021), which formed in 1996. However, a significant problem with the approach taken by both organisations is that their test results are only available to US law enforcement agencies.  Flandrin *et al.* (2014) point out that this decision is contrary to the principle tenet of information sharing in science.  Hence, there is a clear absence of open reproducibility, rendering any results obtained from such tests as non-scientific.

## 3. Research design using the MATEF

The MATEF has been designed to increase levels of trusted practice in the evaluation of malware artefact detection tools. To this end, the framework provides quantitative data collection and analysis capabilities based on the systematic testing of malware analysis tools. As shown in Figure 1, the framework provides a virtualised testing environment that can automatically install the malware analysis tools to be tested, alongside relevant malware samples to generate tool log files (containing details on artefact activity) associated with the execution of the malware. Previously, the use of a virtualised environments for executing malware could present challenges with the malware detecting the use of virtualisation and behaving differently or exiting prematurely. However, Wueest (2014) and Miramirkhani et al. (2017) both argue virtualisation is increasingly pervasive and likely to be an equal source of revenue. Therefore, the use of anti-virtualisation strategies has started to diminish in prevalence.

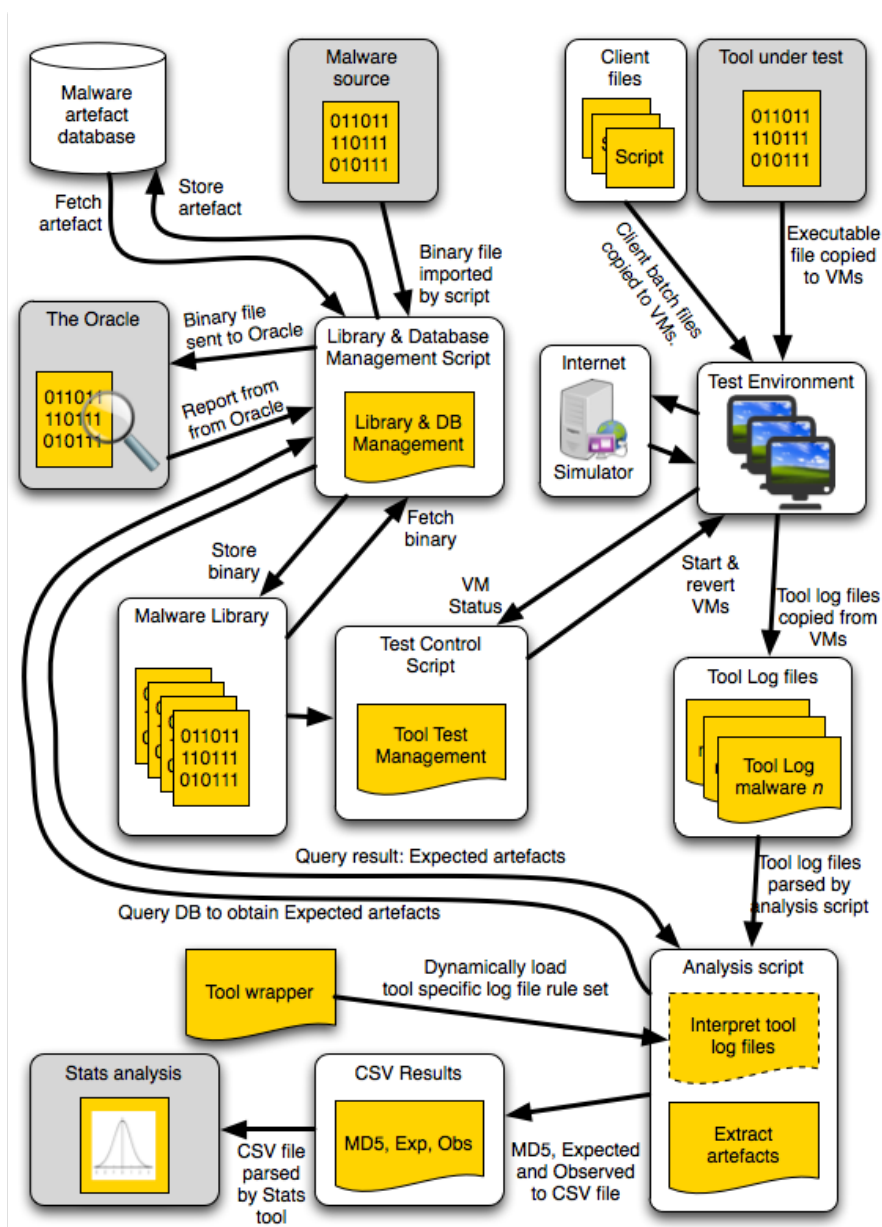

*Figure 1 : MATEF components*

Characterising the research approach supported by the MATEF using the 'Research Onion' model (Figure 2) proposed by Saunders et al. (2007), it can be said to adopt a positivist research philosophy to provide a methodology that was independent of the researcher (and users) and to place the focus of any evaluation on the software tools themselves.

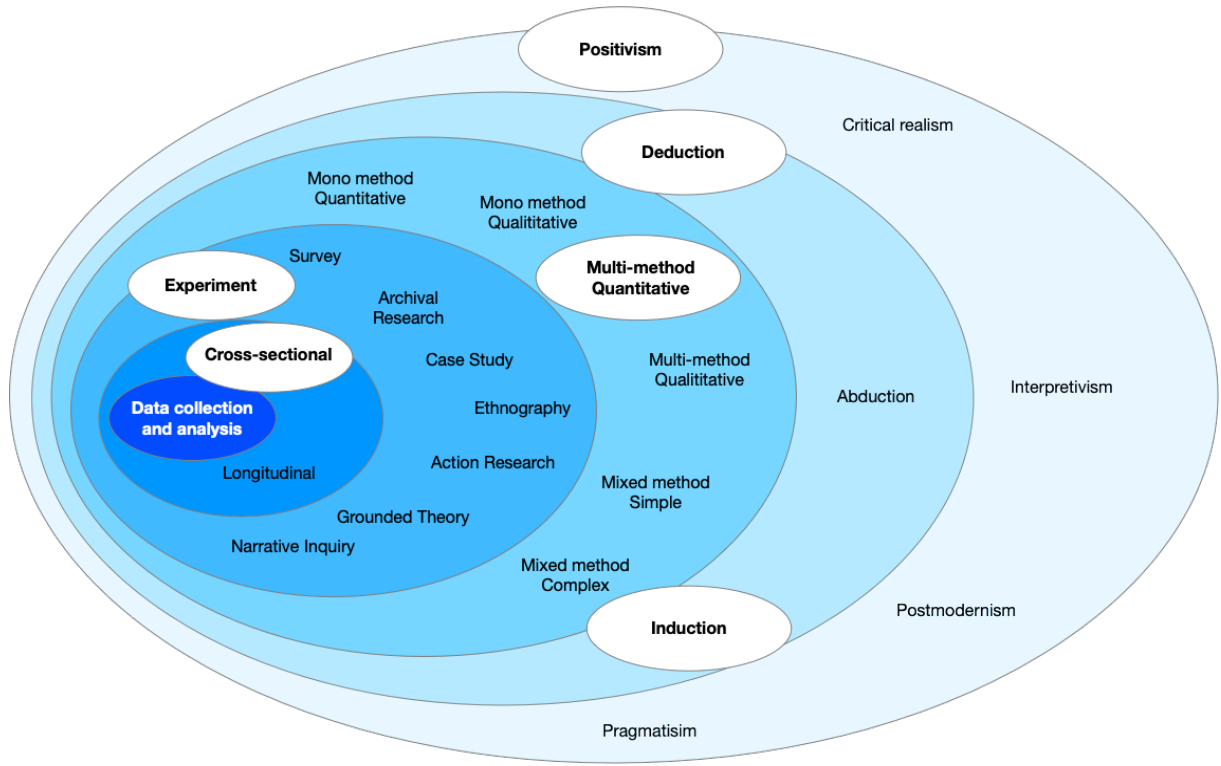

*Figure 2 : Characterising the MATEF using the Research Onion. Image adapted from: Saunders et al. (2007)*

Following the route marked by the white bubbles in Figure 2, we have considered the underlying theoretical framework relevant to a given research question, which could be inductive or deductive. For example, a study of malware epidemiology, could apply the MATEF to a more deductive investigation, calling upon existing epidemiological theory to support generalisation of the findings. On the other hand (as illustrated by the case study presented in this paper), when an investigation of malware analysis tool behaviour has no existing theory to underpin it, MATEF could be used with an inductive approach, with a view to generalising the results using probability analysis.

Irrespective of the approach to theory formulation, the MATEF methodology is based on multi-method quantitative analysis of data gathered through experimentation on the malware analysis tools under investigation. This informs and constrains the types of hypotheses that can be tested using the MATEF to those that involve empirically observable phenomena that can be quantified through measurement. For example, this would include hypotheses based on metrics such as:

- countable artefacts detected by malware analysis tools
- time for which malware executes
- time for which malware analysis tools executes
- number of types of malware that a tool is tested with
- frequency of observable events
- changes in entropy levels in data

Evaluating Malware Forensics Tools. Copyright © 2022 Ian M. Kennedy                5

Metrics such as these can be used to underpin a variety of research questions, such as those listed in Table 1.

| RQ1 | Is tool X a better choice than tool Y for observing artefacts of malware (MW)? |
| --- | --- |
| RQ2 | Which of two tools offering a specific capability is the better choice? |
| RQ3 | Does the type of malware affect the perceived accuracy of a given tool? |
| RQ4 | What is the optimal length of time needed to execute MW when under analysis? |
| RQ5 | What is the variability of the optimal execution time for a tool between MW types? |
| RQ6 | Does one tool have greater levels of repeatability than another? |
| RQ7 | Is one tool more readily detected by malware over another? |
| RQ8 | Does a given tool observe more artefacts OS A compared to OS B? |

*Table 1 : Sample Research Questions*

To clarify, the focus of this work is to demonstrate how the MATEF can be implemented to design a variety of experiments to answer research questions (such as those illustrated in Table 1) framed to evaluate artefact detection tools used within a forensic investigation.

The components of the MATEF (Figure 1) can be configured differently to accommodate different research questions, as needed. For example, to address RQ6 (see Table 1) no oracle component is needed to measure the repeatability of observable artefacts. Also, different test environments can be swapped in and out, such as changing the operating system (e.g., RQ8, Table 1).

In addition to accommodating different research questions, this flexibility in design means the MATEF can accommodate swapping out components which are subject to availability. For example, where a specific Oracle resource is no longer available or a particular source of malware binaries to ingest into the Malware Library becomes unavailable.

Once one or more research questions are identified, hypotheses can be generated to formalise the result of subsequent experiment work. In this work, we have selected RQ4 and RQ1 as examples from the list of sample research questions in Table 1 to demonstrate the implementation and application of the MATEF. These will be introduced in the case study (Section 5). Prior to this, however, we present in the next section a description of the implementation used to support the case study.

## 4. Example implementation the MATEF

To demonstrate how to utilise the MATEF we present a suggested implementation in this section and follow this with a series of experiments in a case study in the next section.

To implement the MATEF each of the components shown in Figure 1 needs to be implemented. The chosen implementation method for the case study, together with a supporting rationale, is provided for each component under the sub-headings that follow.

The Oracle
At the time the MATEF was originally implemented to collect data a variety of currently active online sandboxes were available as an Oracle source. Examples include: Comodo Valkyrie (2022), Joe Sandbox (2017), Cuckoo Sandbox (2022), and IObit (2022). In addition, two sandboxes (available at the time) that are no longer available were Anubis (http://anubis.iseclab.org) and Threat Expert (http://www.threatexpert.com). Of these latter two, Anubis was used to provide the reference dataset for the subsequently analysed case study that follows. The rationale for this was that the Anubis platform provided the simplest mechanism to retrieve the reports and ingest them into the MATEF. Whichever source is used for the Oracle, a report (typically in XML) is generated providing detail on each type of artefact (eg: file, registry keys, etc). Details from the reports are stored in the malware artefact database, implemented here in sqlite (see 'Database platform' below) using the md5 hash as the key. The data stored in the malware artefact database (sourced by the Oracle) is used to calculate what is referred to hereafter as the 'expected number of artefacts'. For the purposes of the case study that follows, this dataset represents an approximated 'ground truth'.

The resilience of the MATEF design is demonstrated by the fact that although the originally used sandbox is no longer available, the Oracle module can be swapped out for a contemporary alternative. It is important to note that the MATEF operates by evaluating tool observations relative to a reference dataset. The absence of absolute 'ground truth' in such a dataset is a limitation of the online sandboxes themselves and not the Oracle component.

Scripting platform
The MATEF has been implemented in Python, due to it being easy to maintain and very well supported in terms of libraries and open-source code. However, this could notionally be any other high-level language capable of controlling VMs, movement of files and access to the chosen database platform.

Virtualisation platform
VMWare (VMWare, 2016) was chosen to enable multiple virtual machines (VMs) to be operated remotely and via a scriptable interface. The primary reason for using VMWare for the case study was it was readily available and supported within the department. The flexibility of this platform also enabled multiple VMs to be instantiated in parallel to expediate testing. For the case study that follows, a total of sixty (60) identical virtual machines (VMs), each running a vulnerable version of Windows, was constructed. Each VM was configured to start, run for a predefined time (determined in the case study), shutdown and reset to a known clean state between each use. The hypervisor (running CentOS) provided a level of isolation between Windows malware and any physical Windows hosts on the physical network. In addition, the network environment of each VM was configured so that each VM was isolated from others running in parallel and from the Internet. Network isolation was also configured to enable remote access though a single gateway endpoint. Internet services were provided through simulation.

Internet simulator
Over 90% of malware uses DNS services when run (Lee et al., 2019). Furthermore, Gilboy (2016) advises the use of Internet services can help defeat some malware defence mechanisms designed to change how the file executes. For the case study that follows, the module was implemented using the INetSim suite (Hungenberg and Eckert, 2020), due to it being easy to use, reasonably well maintained, and well documented.

Database platform
SQLite was chosen as the database platform as it is relatively light weight and easy to configure. The pervasiveness of the database storage format also meant that documentation and example code was readily available to support the implementation of this module into the MATEF.

Malware source

For the purposes of the case study, a single source module linked to a feed provided by the website VirusTotal (2010) was used. An established working relationship with the operator of the site meant this was relatively easy to arrange. For the duration of the project, this yielded more than 350,000 binary files that were ingested into the Malware Library.

Using the implementation of the MATEF described above, the next section demonstrates its application for investigating some example research questions relevant to evaluating malware forensics tools.

## 5. Case study: investigating impact of tool execution time

To demonstrate an implementation of the MATEF a case study is presented that explores the research question cited in the Introduction, namely "Can a systematic basis for trusted practice be established for evaluating malware artefact detection tools used within a forensic investigation?". The case study is not intended to be a full-scale comparison or investigation of this question, but simply a proof-of-concept implementation.

Our approach was to design a series of experiments to examine the impact of the execution time of different analysis tools on their ability to detect malware artefacts. No existing theory was identified and so an inductive approach was taken with a view to generalising the results through probability. The two tools chosen were Process Monitor and TCPVCon, both available from Sysinternals (2020).

### 5.1 Hypothesis formulation

Two experiments are presented as examples of using the MATEF to evaluate tools used to study malware during a digital forensic investigation. The first is designed to address RQ4 (see Table 1):

*RQ4: What is the optimal length of time needed to execute MW when under analysis?*

To answer this question, it is necessary to execute a malware binary for a variety of different durations to examine the impact upon the tool under evaluation. This approach enabled us to consider if there was an optimal execution time, see $H_1$.

| | |
|---|---|
| $H_{1.0}$ | Changing the execution time of malware has no significant impact on the number of malware artefacts observed by a given tool. |
| $H_{1.1}$ | Changing the execution time of malware has a significant impact on the number of malware artefacts observed by a given tool. |

*Table 2 : Hypothesis 1*

The second experiment was designed to address RQ1 (see Table 1):

*RQ1: Is tool X a better choice than tool Y for observing artefacts of malware?*

Evaluating Malware Forensics Tools. Copyright © 2022 Ian M. Kennedy    8

For this question the approach taken was to compare two tools to determine if there was any significant difference between the number of artefacts observed when monitoring the same malware, under the same conditions, see $H_2$.

| | |
|---|---|
| $H_{2.0}$ | There is no significant difference on the number of malware artefacts observed by Tool A when compared to Tool B, under the same conditions. |
| $H_{2.1}$ | Tool A can detect a significantly greater number of artefacts when compared to Tool B, under the same conditions. |
| $H_{2.2}$ | Tool B can detect a significantly greater number of artefacts when compared to Tool A, under the same conditions. |

*Table 3 : Hypothesis 2*

### 5.2 Experiment configuration

**Test Run:** From a library of over 350,000 malware binaries, a sample of 4,800 files known to produce network artefacts were selected initially at random, identified as 'Test ID 144', see Table 4. Refer to 'Test Run Group' (below) regarding the selection of data for subsequent runs. The sample was then divided equally amongst the 60 available VMs, providing an in-tray of 80 malware binaries to be analysed by each VM. Each VM was started in parallel and then had the tool under analysis copied into each of the environments. Each VM was then populated with a single malware binary from one of the 80 malware binaries allocated to it. At the conclusion of the test, the log file produce by the tool was extracted from each VM for subsequent analysis. Each VM was then shut down and reset, before commencing the next Test Run, where the next malware binary in each VM's corresponding in-tray was used. This collective execution of multiple VMs (in this case 60) is referred to hereafter as a 'Test Run' and is represented as a single row in Table 4 and Table 5.

**Test Run Group:** The Test Run was executed three times using the same malware binaries as used for the initial Test Run (Test ID 114). All other conditions were kept the same, creating a group, referred to hereafter as the 'Test Run Group'. This subsequently allowed malware binaries that produced highly variable numbers of artefacts to be identified by hash and filtered out. The rationale for this decision was to isolate variations in the number of artefacts observed as a result of malware behaviour, as opposed the variations due to the tool under test. As $H_1$ and $H_2$ collectively consider multiple tools and execution times, each Test Run Group changed one of these two control variables, see Table 4 and Table 5.

**Dataset generation:** Each tool tested was configured to monitor all the artefacts it could observe to maximise the data captured, recording the observations in log files. The MATEF enables datasets of different observed artefact types to be generated from these source log files.

For example, a dataset containing only open ports observed or files created can be generated for analysis. When created, the dataset contains three fields: the MD5 Hash of the malware binary, the expected number of artefacts and the observed number of artefacts. Any combination of the following artefact datatypes can be chosen for inclusion in the datasets produced: File or Mutex events, Registry events, Port events, RPort events, File only events, or Mutex only events.

Using port activity as an example artefact type, two tools were selected for comparison. For each tool, four different execution times (10 seconds, 1 minute, 5 minutes and 10 minutes) were chosen to explore $H_1$. A Test Run Group was created for each of these conditions, meaning each Test Run was repeated three times, as described above under Test Run Group. This resulted in 12 datasets for each tool under test, published on the Open Research Data Online repository (Kennedy, 2022) and summarised in Table 4 and Table 5.

| Test ID | Tool | Time | Dataset | Number Log files | Dataset ID | Subset Dataset (Repeatable) | Number of Rows |
|---|---|---|---|---|---|---|---|
| 114 | ProcMon | 1 min | Random | 3830 | A.1 | Process Monitor (1 minute) | 2803 |
| 126 | ProcMon | 1 min | 114 | 3554 | | | |
| 127 | ProcMon | 1 min | 114 | 3632 | | | |
| 117 | ProcMon | 5 min | 114 | 3234 | A.2 | Process Monitor (5 minutes) | 1056 |
| 134 | ProcMon | 5 min | 114 | 2495 | | | |
| 135 | ProcMon | 5 min | 114 | 2399 | | | |
| 118 | ProcMon | 10 min | 114 | 1881 | A.3 | Process Monitor (10 minutes) | 1360 |
| 136 | ProcMon | 10 min | 114 | 1647 | | | |
| 137 | ProcMon | 10 min | 114 | 1588 | | | |
| 147 | ProcMon | 10 sec | 114 | 3435 | A.4 | Process Monitor (10 seconds) | 416 |
| 148 | ProcMon | 10 sec | 114 | 3416 | | | |
| 149 | ProcMon | 10 sec | 114 | 2917 | | | |

*Table 4 : Process Monitor datasets*

| Test ID | Tool | Time | Dataset | Number Logfiles | Dataset ID | Subset Dataset (Repeatable) | Number Rows |
|---|---|---|---|---|---|---|---|
| 115 | TCPVCon | 1 min | 114 | 3380 | B.1 | TCPVCon (1 minute) | 1259 |
| 128 | TCPVCon | 1 min | 114 | 3741 | | | |
| 129 | TCPVCon | 1 min | 114 | 3217 | | | |
| 116 | TCPVCon | 5 min | 114 | 3984 | B.2 | TCPVCon (5 minutes) | 1478 |
| 130 | TCPVCon | 5 min | 114 | 1656 | | | |
| 131 | TCPVCon | 5 min | 114 | 1944 | | | |
| 119 | TCPVCon | 10 min | 114 | 4021 | B.3 | TCPVCon (10 minutes) | 1298 |
| 132 | TCPVCon | 10 min | 114 | 2230 | | | |
| 133 | TCPVCon | 10 min | 114 | 1677 | | | |
| 150 | TCPVCon | 10 sec | 114 | 4006 | B.4 | TCPVCon (10 seconds) | 632 |
| 151 | TCPVCon | 10 sec | 114 | 3409 | | | |
| 152 | TCPVCon | 10 sec | 114 | 3129 | | | |

*Table 5 : TCPVCon datasets*

**Filtering for repeatability:** The initial Test Run (Test ID 114 in Table 4) selected 4,800 binary files at random, known to produce network artefacts. Two issues with repeatability were noted: The first being that not all malware binary files in the sample executed successfully or for all three runs in a Test Run Group. For example, Test ID 114 selected at random 4,800 mw binaries known to produce network artefacts. Of these 4,800 a total of 3,380 binaries ran successfully and produced corresponding log files (see Table 4). The same 4,800 binaries were executed a second time (Test126), resulting in 3,554 successful executions and corresponding log files. The same 4,800 binaries were then executed for a third time (Test127), resulting in 3,632 successful executions and corresponding log files. The secondly issue was that from the subset of those malware binaries that did execute three times, the same number of artefacts was not always observed on all three runs. It is for this reason that the number of log files produced is neither 4,800 nor consistent, even within a Test Run Group.

Collectively, these three tests operating on the same data are referred to as a 'Test Run Group'. Out of the three datasets, a total of 2,803 binaries successfully executed for all three tests and produced the same number of observable artefacts for each test. For this reason, these 2,803 binaries were considered repeatable and formed the Dataset ID A.1.

Using the same 4,800 selected binaries, the same methodology was used to generate the remaining datasets A.2-4 and B.1-4. This created a total of eight datasets (four per tool under test), see the 'Subset Dataset (Repeatable)' and 'Number Rows' columns in Table 4 and Table 5.

With the repeatable datasets produced for both Process Monitor and TCPVCon, the next stage was to consider the analysis strategy.

## 5.3 Analysis strategy

In this section we present our strategy for identifying and selecting the metrics that could be used to address the hypotheses (see Table 2 and Table 3). Following this, a brief overview of how the datasets were combined to address each hypothesis is given. Finally, a rationale on the choice of statistical tests used is presented.

**Identifying what to measure:** To maximise the information that can be extracted from the datasets listed in Table 4 and Table 5, Stevens' classification typology commonly known as 'Levels of Measurement' (Stevens, 1946) was applied to the data, see Table 6. The most fundamental level is shown on the bottom row of Table 6. Moving up the table, each level inherits the properties of the preceding, meaning it can accommodate the types of data of the levels below it. Furthermore, each level has associated with it several valid operations, see Table 6.

| Level | Examples | Operations |
|---|---|---|
| Ratio | File size of malware | =, <>, <, >, +, -, *, / |
| Interval | Number of ports opened by malware | =, <>, <, >, +, - |
| Ordinal | Threat level of malware (eg: Low, Medium, High) | =, <>, <, > |
| Nominal | Port number, Filename, Registry key name | =, <> |

*Table 6 : Measurement levels*

Under this scheme, the port number of each opened port on a computer and the name of a file or registry key created are all examples of *nominal* data. The names and the number (in the case of port numbers) are nothing more than labels that refer to the artefact it represents. There is, for example, no inherent difference between the network port 80 and port 443. Although the former is commonly used for unencrypted web browser traffic and the latter for encrypted traffic, there is little else that can be determined from comparing them. To state that one is greater than the other is or there is a 'difference' of 363 between them is meaningless, as the port numbers do not represent a quantity.

Both Hypothesis 1 (Table 2) and Hypothesis 2 (Table 3), examine 'the number of malware artefacts' to determine an outcome. Hence, although the individual malware artefacts are nominal in nature, a count of their numbers represents a quantity. Although individual artefacts are nominal in nature, the operations that can be applied to such data is limited. However, analysis of their quantity provides scope for extracting a greater level of information. Quantity data can be compared not just in an ordinal fashion (eg: one binary produces more artefacts than another) but also in terms of how much they differ (eg: one binary produces 10 more artefacts than another).

This approach enables a greater number of operations associated with interval data to be applied to the data, see Table 6. However, although differences on a scale can be meaningful and comparable to other differences on the same scale, the zero point on the scale is arbitrary (Panik, 2005); hence ratios of interval scale values are meaningless. In practice, this means comparing tools on nominal data (such as the number of detected open ports) would not enable a conclusion such as 'Tool A is twice as good as Tool B' to be reached.

The information extracted from the observations was extended further by comparing the quantity of observed artefacts with the expected number of artefacts produced for a given malware binary, provided by the MATEF Oracle, see section 4. The comparison of a pair of observed values (one for each execution time, for Hypothesis 1) would only provide a measure of how similar the distribution of observations are at different lengths of execution. By also including the expected number of artefacts for each data point and calculating the absolute difference of the corresponding observed value from this, it becomes possible to gain a measure of the error in each observation.

Note that 'error' in this context is defined as the difference between an estimated 'ground truth' (as provided by the Oracle) and the observed value. Plotting the frequency of these errors then produces a distribution of the absolute error of the observations between two execution times.

**Testing strategy for $H_1$:** To consider $H_1$, datasets with different execution times from the same tool were compared to determine if there is a significant difference on the number of artefacts observed for different execution times of a tool. To create paired values for each malware binary, analysis was limited to those binaries observed in both datasets (ie: two different execution times). Datasets were combined by hash and each tool was subjected to three comparative analyses after outliers were removed. Two different tools were considered: Process Monitor (Table 7) and TCPVCon (Table 8).

| Process Monitor | | | |
|---|---|---|---|
| Test | Analysis description | Datasets | Number rows |
| 1.1 | Comparing 1 minute to 10 second execution times | A.1, A.4 | 333 |
| 1.2 | Comparing 1 minute to 5 minute execution times | A.1, A.2 | 829 |
| 1.3 | Comparing 1 minute to 10 minute execution times | A.1, A.3 | 1056 |

*Table 7 : Comparing Process Monitor for different execution times ($H_1$)*

| TCPVCon | | | |
|---|---|---|---|
| Test | Analysis description | Datasets | Number rows |
| 2.1 | Comparing 1 minute to 10 second execution times | B.1, B.4 | 274 |
| 2.2 | Comparing 1 minute to 5 minute execution times | B.1, B.2 | 675 |
| 2.3 | Comparing 1 minute to 10 minute execution times | B.1, B.3 | 569 |

*Table 8 : Comparing TCPVCon for different execution times ($H_1$)*

**Testing strategy for $H_2$:** For the second hypothesis, datasets with the same execution times from different tools were compared to determine if there is a significant difference on the number of artefacts observed between tools for comparable execution times of each tool. As noted under 'Dataset generation' (§5.2), the initial dataset generated from TestID 114 was a random selection of 4,800 binary files, known to produce network artefacts. Hence, neither tool was assumed to be targeted by any malware. As before, datasets were combined by hash and outliers were removed, see Table 9.

Having organised the data into a form to address each hypothesis, consideration must be given to the appropriate choice of statistical tests for analysis.

| Test | Analysis description | Datasets | Number rows |
|---|---|---|---|
| 3.1 | Process Monitor vs TCPVCon, run for 10 seconds | A.1, B.1 | 125 |
| 3.2 | Process Monitor vs TCPVCon, run for 1 minute | A.2, B.2 | 994 |
| 3.3 | Process Monitor vs TCPVCon, run for 5 minutes | A.3, B.3 | 496 |
| 3.4 | Process Monitor vs TCPVCon, run for 10 minutes | A.4, B.4 | 554 |

*Table 9 : Comparing Process Monitor and TCPVCon for corresponding execution times ($H_2$)*

**Selection of statistical test:** For each of the eight datasets (A.1-4 and B.1-4) an absolute error value (|Expected number of artefacts – Observed number of artefacts|) was calculated. From the resulting distributions, checks for normality were applied using the Kolmogorov-Smirnoff (Sheskin, 2011, p. 261) and Shapiro-Wilk (Sheskin, 2011, p. 240). In all cases, the distributions did not follow a Normal distribution.

The nonparametric Wilcoxon Signed Rank test (Sheskin, 2011, p. 809) was selected to perform the analysis. This is because it is an established and appropriate statistical test for comparing distributions containing paired observations that are not normally distributed; or where one or more of the assumptions for the equivalent *t test* are saliently violated. The test accepts either ordinal or interval data. Competing tests such as the *binomial sign* test (Sheskin, 2011, p. 823) and *McNemar* test (Sheskin, 2011, p. 835) were discounted, as they either had less statistical power or only operate on nominal data.

With the strategy in place for analysis, the next step was to perform the analysis and examine the results.

Evaluating Malware Forensics Tools. Copyright © 2022 Ian M. Kennedy                    13

## 5.4 Results

Considering H$_1$, and using Test 1.1 as an example, the two absolute error values were supplied to a Wilcoxon Signed Rank test, using SPSS v21. This produced a rejection of the Null Hypothesis (H$_{1.0}$):

**Hypothesis Test Summary**

| Null Hypothesis | Test | Significance | Decision |
|---|---|---|---|
| The median differences between ProcessMon_1min and ProcessMon_10sec equals 0 | Related samples Wilcoxon Signed Rank Test | 0.011 | Reject the null hypothesis |
| Asymptotic significances are displayed. The significance level is 0.05 | | | |

| | |
|---|---|
| Total N | 333 |
| Test Statistic | 28.000 |
| Standard Error | 5.534 |
| Standardised Test Statistic | 2.530 |
| Asymptotic Significance (2-sided test) | 0.011 |

The effect size ($r$) is given by:

$$r = \frac{z}{\sqrt{N}} = \frac{2.530}{\sqrt{333}} = 0.1386$$

where $z$ is the Standardised Test Statistic and $N$ is the number of rows in the dataset.

This means that for the *Process Monitor* tool, the differences between the expected and observed number of ports opened during a 1 minute execution time (Median=427) were significantly different to the differences between the expected and observed number of ports opened during a 10 second execution time (Median=427), Test Statistic ($T$) = 28, $p$ = 0.011, $r$ = 0.1386.

The same analysis was applied to the remaining tests (*mutatis mutandis*) for H$_1$, see Table 10:

| Results for Process Monitor | | | | | | | | |
|---|---|---|---|---|---|---|---|---|
| Test | Description | $r$ | $z$ | SE | $T$ | $p$ | $N$ | Result |
| 1.1 | 1m vs 10s | 0.1386 | 2.530 | 5.534 | 28.000 | 0.011 | 333 | Reject H$_1$ |
| 1.2 | 1m vs 5m | 0.0634 | 1.826 | 2.739 | 10.000 | 0.068 | 829 | Retain H$_1$ |
| 1.3 | 1m vs 10m | 0.0418 | 1.357 | 5.895 | 22.000 | 0.175 | 1056 | Retain H$_1$ |

Table 10 : Results for H$_1$ (Process Monitor)

Referring to Table 10, for increases in the execution time a reduction in effect size ($r$) was observed, indicating it becomes increasingly more difficult to detect any differences between the two distributions being compared in each test. This is reflected in the value of $z$ (in effect a 'signal to noise ratio') that reduces as the execution time increases. However, only Test 1.1 can reject H$_1$, suggesting that the optimal execution time to observe malware when using Process Monitor is between 10 seconds and 1 minute. Executing Process monitor for more than 1 minute did not produce any perceived benefit in terms of the number of artefacts observed by the tool.

The same methodology was applied to the *TCPVCon* tool (*mutatis mutandis*) for $H_1$, see Table 11:

| Test | Description | *r* | *z* | *SE* | *T* | *p* | *N* | Result |
|---|---|---|---|---|---|---|---|---|
| 2.1 | 1m vs 10s | 0.0811 | 1.342 | 1.118 | 3.000 | 0.180 | 274 | Retain $H_1$ |
| 2.2 | 1m vs 5m | -0.0172 | -0.447 | 1.118 | 1.000 | 0.655 | 675 | Retain $H_1$ |
| 2.3 | 1m vs 10m | NaN | NaN | 0.000 | 0.000 | 1.000 | 569 | Retain $H_1$ |

Table 11 : Results for $H_1$ (TCPVCon)

As observed with Process Monitor (Table 10), the results for TCPVCon (Table 11) indicate a reduction in effect size (*r*) was observed (to the point it is undetectable) as the execution time is increased. Furthermore, none of the execution times resulted in retaining $H_1$. This means no effect was observed as a result of running TCPVCon for different durations of time.

Test 2.3 (Table 11) produced a Standard Error (*SE*) value of zero. The impact of this means that the Standardised Test Statistic (*z*) cannot be calculated and thus the effect size cannot be determined. Furthermore, an *SE* value of zero indicates the median of the differences between the two distributions (the 1 minute and 10 minute execution times) is also zero, i.e.: there is no change between the two distributions.

Considering $H_2$, the same methodology to compare the two distributions was applied (*mutatis mutandis*) to compare Process Monitor (PM) and TCPVCon (TCPVC) for comparable execution times, see Table 12.

The results from Table 12 indicate there is no significant difference in the number of artefacts observed when executing Process Monitor or TCPVCon under the same test conditions.

| Test | Description | *r* | *z* | *SE* | *T* | *p* | *N* | Result |
|---|---|---|---|---|---|---|---|---|
| 3.1 | PM-TCPV (10s) | NaN | NaN | 0.000 | 0.000 | 1.000 | 125 | Retain $H_2$ |
| 3.2 | PM-TCPV (1 min) | -0.0605 | -1.908 | 27.253 | 63.500 | 0.056 | 994 | Retain $H_2$ |
| 3.3 | PM-TCPV (5 mins) | 0.0635 | 1.414 | 1.061 | 3.000 | 0.157 | 496 | Retain $H_2$ |
| 3.4 | PM-TCPV (10 mins) | -0.0425 | -1.000 | 0.500 | 0.000 | 0.317 | 554 | Retain $H_2$ |

Table 12 : Results for $H_2$ (Process Monitor vs TCPVCon)

# 6. Evaluation

Applying the MATEF to the case study has demonstrated it is possible to address practical questions, such as determining the optimum execution time to use when evaluating an unfamiliar software tool. It is important to note that many of the sample questions in Table 1 (including the two selected for the case study) are comparative in nature. Hence the need for knowledge of absolute ground truth is mitigated. However, although not a limitation of the MATEF, the lack of 'ground truth' data has meant we have only been able to determine a relative (as opposed to absolute) accuracy for individual tools. Furthermore, although not an issue with the MATEF itself, it is acknowledged that the case study is relatively small in scale, involving only two tools across ten different experiments to exemplify the implementation of the framework.

The implementation of MATEF did pose some challenges with performance. Initially, we had multiple VMs instantiated in simultaneously to expedite testing. However, this created a resource bottleneck on the VM Manager, so VMs that were intended to be run in parallel were instantiated with a short (configurable) offset of 10 seconds between them to distribute the load on the system.

A related issue was the time it took to perform a single test, where each test had approximately 10 minutes of overhead. To illustrate this, a 10 second test took just under 10 minutes to complete the file preparation, instantiating, testing, reverting and log file capture steps. Similar overheads were observed for tests of other durations.

The case study implementation of the MATEF excluded malware that was not repeatable (see §5.2, Table 4 and Table 5) to minimise the impact on artefacts observed by software tool undergoing evaluation. Whilst this might be considered to reduce the representativeness of the malware used to test the tools, it is not a limitation of the MATEF, but instead a decision taken for the case study. An improved methodology to handle such variations in malware or even where a different research question is the focus of the study could address this issue.

It is also possible that the results for a given tool will vary between different organisations using the MATEF. This is not an uncommon problem and has been identified in a conventional computer forensics context by Garfinkle *et al.* (2009). It is also recognised by the VIM standard (JCGM, 2008), which defines this situation in terms of reproducibility. Thus, rather than being a 'problem', this phenomenon is considered a useful by-product of the framework that would facilitate any future cross-lab study into reproducibility of tools.

## 7. Conclusions

This paper demonstrates how the MATEF provides a platform upon which to design experiments to evaluate the performance of malware analysis tools. To this end, our case study illustrates the use of MATEF to address two research questions, selected from a bank of sample questions presented in Table 1.

In response to the first research question, the results tentatively indicate the choice of tool can have an impact on the number of artefacts observed under different execution times. This indicates it is possible to identify an optimal duration for which to run malware binaries under analysis. Further testing with additional software tools would be beneficial to validate these findings.

With regards to the second research question, comparing the number of malware artefacts observed by two tools running independently under the same conditions, showed there was no significant difference between them. Hence, once the operating conditions were set, there was no perceived penalty in choosing one tool over another. Again, further testing with additional software tools would be beneficial to validate these findings.

The application of the MATEF has enabled us to demonstrate it is possible to apply a systematic approach to comparing tools and their use during a forensic investigation involving malware. This reduces uncertainty and so enhances the level of trust placed in malware forensic practice.

## 8. Further work

Although not a limitation of the MATEF itself, the availability of absolute ground truth data would enhance the range of studies that could be performed by the framework. In terms of implementation, performance issues could be addressed through use of compiled code, better resource management, including optimisation of the hypervisor in use. Further hardening of the virtual environment to minimise detection and evasive behaviour exhibited by malware could be undertaken. Similarly, support for less repeatable malware to enable a wider range of research questions to be addressed could be applied.

Looking beyond tool evaluation, the MATEF could be extended to improve our understanding of the decisions made by practitioners during malware forensic investigations. For example, interpretivist-based experiments that focus more on human interaction and less on automation could be developed using the MATEF to monitor which tools are used when presented with different types of malware. Such studies could be cross-sectional, or more longitudinal, to gain deeper insights into practitioner actions over time.

## Funding

This work was partly supported by the Engineering and Physical Sciences Research Council grant, SAUSE: Secure, Adaptive, Usable Software Engineering (EP/R013144/1).